\documentclass[a4paper,10pt,DIV=12,flalign, abstracton]{scrartcl}

\usepackage[utf8]{inputenc}
\usepackage{graphicx}
\usepackage{import}
\usepackage{color}
\usepackage[numbers,sort&compress]{natbib}
\usepackage{amsmath}
\usepackage{amsfonts}
\usepackage{amssymb}
\usepackage{caption}
\usepackage{subcaption}
\usepackage{csquotes}
\usepackage{mhchem}
\usepackage{multirow}
\usepackage{tabularx}
\usepackage[noblocks]{authblk} 

\title{On the question of the sign of size effects in the elastic behavior of foams}
\author[1]{Stephan Kirchhof} 
\author[1]{Alfons Ams}  
\author[1,2]{Geralf Hütter} 
\affil[1]{Technische Universität Bergakademie Freiberg, Institute of Mechanics and Fluid Dynamics, Freiberg, Germany}
\affil[2]{Brandenburg University of Technology Cottbus-Senftenberg, Chair of Mechanics and Numerical Methods, Cottbus, Germany}

\date{\today}

\newcommand{\Eapp}{E_{\mathrm{app}}}
\newcommand{\Eapplong}{E_{\mathrm{app,l}}}
\newcommand{\Eappbend}{E_{\mathrm{app,b}}}
\newcommand{\Eeff}{E_{\mathrm{eff}}}
\newcommand{\Eeffbx}{E_{\mathrm{eff-b,x}}}
\newcommand{\Eeffby}{E_{\mathrm{eff-b,y}}}
\newcommand{\Eeffl}{E_{\mathrm{eff-l}}}
\newcommand{\Ebulk}{E_{\mathrm{bulk}}}
\newcommand{\Gapp}{G_{\mathrm{app}}}
\newcommand{\Geff}{G_{\mathrm{eff}}}

\newcommand{\sizeeffratio}{\Omega}
\newcommand{\sizeeffratiobend}{\sizeeffratio_{\mathrm{b}}}
\newcommand{\sizeeffratiolong}{\sizeeffratio_{\mathrm{l}}}
\newcommand{\sizeeffratiotors}{\sizeeffratio_{\mathrm{t}}}

\newcommand{\rhobulk}{\varrho_{\mathrm{bulk}}}
\newcommand{\nubulk}{\nu_{\mathrm{bulk}}}
\newcommand{\nueff}{\nu_{\mathrm{eff}}}
\newcommand{\mspec}{m_\mathrm{specimen}}
\newcommand{\lspec}{\ell}
\newcommand{\rhorel}{\varrho_\mathrm{rel}}

\newcommand{\eigenfreqangbend}{\omega_{\mathrm{b}}}
\newcommand{\eigenfreqangbendx}{\omega_{\mathrm{b}x}}
\newcommand{\eigenfreqangbendy}{\omega_{\mathrm{b}y}}
\newcommand{\eigenfreqangtors}{\omega_{\mathrm{t}}}
\newcommand{\eigenfreqanglong}{\omega_{\mathrm{l}}}

\newcommand{\dpx}{d_\mathrm{p-x}}   
\newcommand{\dpy}{d_\mathrm{p-y}}
\newcommand{\dpz}{d_\mathrm{p-z}}
\newcommand{\dpball}{d_\mathrm{p-ball}}

\begin{document}

\maketitle

\begin{abstract}
Due to their good ratio of stiffness and strength to weight, foam materials find use in lightweight engineering.
Though, in many applications like structural bending or tension, the scale separation between macroscopic structure and the foam's mesostructure like cells size, is relatively weak and the mechanical properties of the foam appear to be size dependent.
Positive as well as negative size effects have been observed for certain basic tests of foams, i.e., the material appears either to be more compliant or stiffer than would be expected from larger specimens.
Performing tests with sufficiently small specimens is challenging as any disturbances from damage of cell walls during sample preparation or from loading devices must be avoided. Correspondingly, the number of respective data in literature is relatively low and the results are partly contradictory.

In order to avoid the problems from sample preparation or bearings, the present study employs virtual tests with CT data of real medium-density ceramic foams. A number of samples of different size is \enquote{cut} from the resulting voxel data. Subsequently, the apparent elastic properties of each virtual sample are \enquote{measured} directly by a free vibrational analysis using finite cell method, thereby avoiding any disturbances from load application or bearings.
The results exhibit a large scatter of the apparent moduli per sample size, but with a clear negative size effect in all investigated basic modes of deformation (bending, torsion, uniaxial).
Finally, the results are compared qualitatively and quantitatively to available experimental data from literature, yielding common trends as well as open questions.

\emph{Foam, cellular material, elastic size effect, bending, torsion, virtual testing}
\end{abstract}

\section{Introduction}

Natural and man-made foam materials combine a low weight with remarkable mechanical properties \citep{GibsonAshby}. They can be classified according to several criteria, e.g.\ into open-cell or closed cell foams, whether the bulk material is metallic, ceramic or biological, among others. 
In the last decade, the huge progress in additive manufacturing offered the possibility to design any detail of the meso-structure of foams and thus to tailor specific properties of such foam-like materials, which are nowadays known as lattice materials or metamaterials, depending on particular details of the mesostructure, cf.\ \citep{Bargmann2018, Lakes2020}. In the following, the general term \emph{cellular solids} \citep{GibsonAshby} will be used.

Using cellular solids in engineering applications requires to characterize their \emph{effective mechanical behavior} in order to predict the reliability of components within the design cycle of structures. 
Though, in many application cases (as well as in biological structures), the (apparent) effective mechanical properties like Young's modulus or strength depend on the specimen size. Such so-called size effects \citep{Lakes2020} can be positive or negative, i.e., small specimens may be stiffer or more compliant than expected from the tests of sufficiently large specimens. 
The reason for the appearance of size effects is that the definition of effective properties requires that the structural length scale is considerably larger than the relevant material length scale -- a condition that can be violated by small samples of cellular solids with their relatively large cells.

Experimental evidence of size effects in bending and torsion of foams has been brought by the pioneering works of Lakes and co-workers for bone \citep{Yang1982,Lakes1995a} and a number of polymer foams \citep{Lakes1983,Lakes1986,Anderson1994a,Lakes2015,Rueger2016,Rueger-Lakes2019}.
All these experiments showed a \emph{positive size effect} (\enquote{smaller is stiffer}). In contrast, a \emph{negative size effect} in bending of conventional foams was reported by \citet{Brezny1990} for static loading as well as by Liebold and Müller \citep{Liebold2015,Liebold2016a} for the first bending eigenfrequency.

For meta-materials, a positive size effect in bending has been reported also by \citet{Dunn2019} as well as by \citet{Waseem2013} and Rueger and Lakes \citep{Rueger2017,Rueger2017a,Rueger2018,Rueger2019}. Though, this effect in regular mesostructures can be explained by the shift of material away from the neutral axis \citep{Yoder2019a,Nourmohammadi2020, Wheel2015} according to Steiner's theorem. And consequently, the opposite size effect appears if the surface is realized in another way within the same metamaterial \citep{Frame2018,Ameen2018}.
Recently, \citet{Reasa2020} formulated the hypothesis that there might be different regimes regarding specimen size relative to the cell size.

A negative size effect was also observed in the elastic regime under uniaxial compression by different research groups \citep{Maheo2013,Andrews2001,Rueger2016}.
In contrast, \citet{Chen2002} observed a positive size effect in the yield stress \enquote{constrained compression}. A positive size effect in the plastic behavior (though with stress concentration due to a hole) has also been reported by \citet{Dillard2006}.

Compared to these few experimental data, there is a large number of numerical studies on size effects in foams and meta-materials. Though, their significance for the understanding of the behavior of real foams is currently limited: 
Direct numerical simulation (DNS), i.e., simulations with discretely resolved mesostructure, have mostly been performed for periodic structures \citep{Ha2016,Dunn2019,Wheel2015,Pham2021,Yoder2019a,Ameen2018,Iltchev2015,Glaesener2019,Rizzi2019a,Janicke2013,Huetter2019,Rosi2019}. The few studies on (more or less) irregular mesostructures \citep{Tekoglu2011,Liebenstein2018,Muehlich2020} are limited to plane models (and so are very most of the aforementioned DNS). And even for a stochastically generated 3d model like in \citep{Rajput2019}, the question on the transferability to real foams arises.

In summary, it can be said that there are numerous numerical and experimental studies on size effects in meta-materials with regular mesostructure which proved a positive size in bending or torsion. For conventional foams with stochastic mesostructure, there are only few (and highly idealized) direct numerical simulations available and even less experimental results. These few experimental results are contradictory with respect to the qualitative behavior in bending, i.e., whether there is a positive size effect or a negative one. 
The problem with respective experiments is that the preparation of small specimens and the load application to such specimens is prone to many potential sources of error, in particular from damage of the surface layer due to machining (resulting in artificial negative size effects)  or clamping and load application (resulting in artificial positive or negative size effects) \citep{Brezny1990,Anderson1994a,Rajput2019}.

The scope of the present study is to contribute a profound and statistically validated data set on the size effects of a real foam \emph{eliminating} the aforementioned sources of error. For this purpose, a voxel representation of a real foam is generated using computed tomography (CT) from which specimens of different size and different location are \enquote{cut}, before they are \enquote{tested} virtually by computer simulations.

\section{Material and Methods}

\subsection{Foam Material}
\label{sec:foammaterial}
\begin{figure}
    \centering
    \scriptsize
    \begin{subfigure}[b]{0.48\textwidth}
    	\def\svgwidth{\textwidth}
    	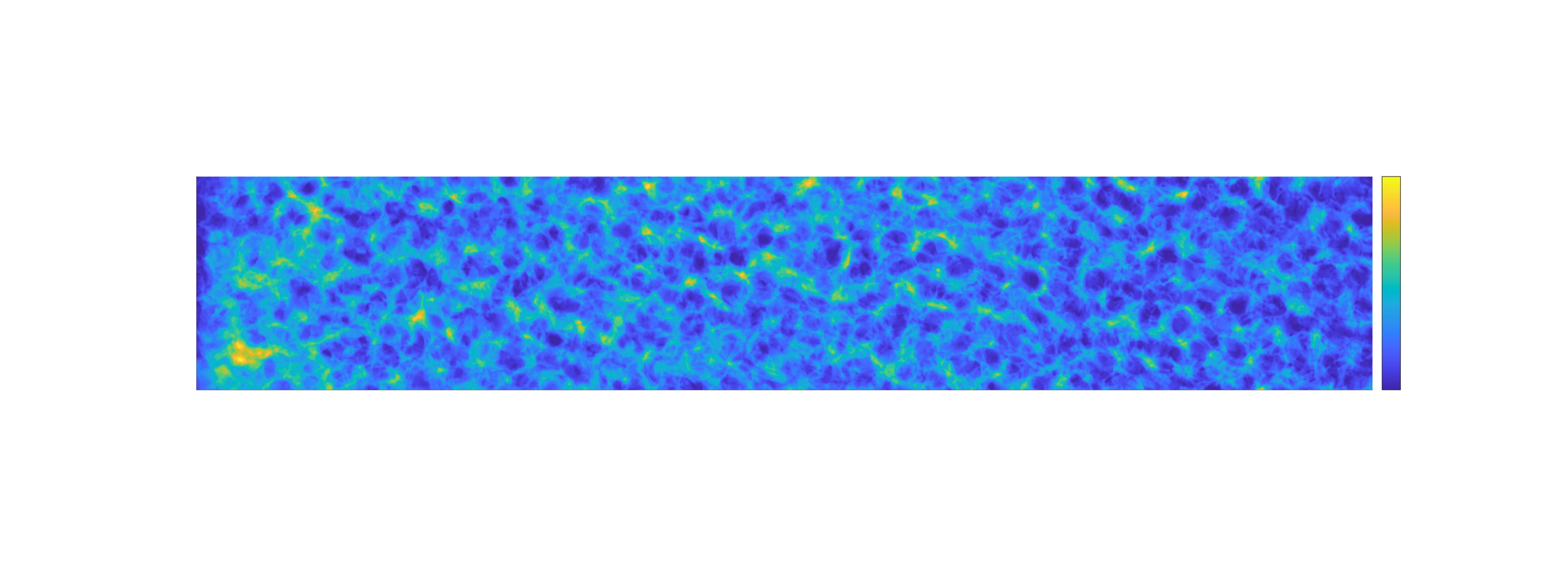
		\caption{specimen 1}\end{subfigure}\hfill
       \begin{subfigure}[b]{0.48\textwidth}
    	\def\svgwidth{\textwidth}
    	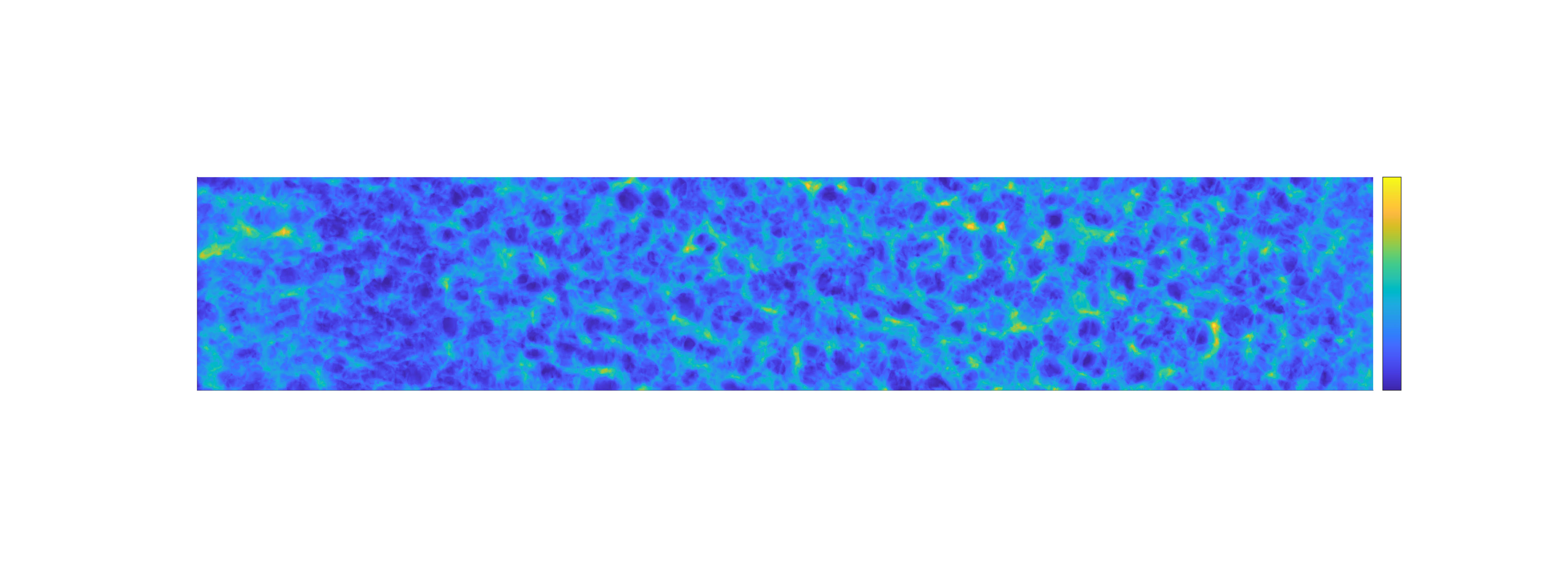
		\caption{specimen 2}\end{subfigure}
    \caption{mean amount of material over depth (y-direction) at certain positions x and z which shows the heterogenity of the foams. Blue areas represent low amounts of material (e.g. holes if value is 0), while yellow areas show massive parts.}
    \label{fig:material_distribution_specimen}
\end{figure}

The specimens in this study are made from ceramic foams of carbon bonded alumina (\ce{Al2O3-C}) with a nominal pore density of 10 pores per inch (ppi). These foams are produced by infiltrating a polyurethane foam by a ceramic slurry \cite{Schwartzwalder.1963}. The slurry is composed of alumina (66 wt.\%), Carbores P as binder (20 wt.\%), Carbon black N991 (6.3 wt.\%) and Graphite AF (7.7 wt\%). Following the infiltration process the foam is centrifuged and fired which results in a incineration of the plastic parts and therefore hollow struts. The process results in 50 specimens, which have a dimension of 30~mm$\times$40~mm$\times$175~mm (W$\times$H$\times$L). Their relative density varies between 16.5\% and 25.7\%. The specimens selected for this study show a relative density of 19.4\% (specimen 1) and 20\% (specimen 2). The mean amount of material over thickness dimension is shown in figure~\ref{fig:material_distribution_specimen} to provide an impression of the heterogeneity of the material.

Furthermore, the mechanical properties of the strut material are of particular interest. From tests with the bulk material the Young's modulus of the ceramic bulk material is determined as $\Ebulk=20\,\text{GPa}$ while the Poisson's ratio can be assumed as $\nubulk=0.2$ \cite{Werner.2014}. The density of the bulk carbon bonded alumina is estimated via volume method with a gas pycnometer as $\rhobulk=2990.6\,\text{kg/m$^3$}$.

The advantage of using the ceramic foam over the original polyurethane foam lies in the larger strut diameters, which reduce the effort required for the CT scans. Larger struts allow for usage of bigger voxel dimensions and therefore larger scan volumes following the measurement principle and the intercept theorem in CT. Additionally, larger strut diameters allow for larger element sizes in direct numerical simulation which reduces the computational effort.

\subsection{Generation of virtual representations by computed tomography}

To obtain volume images of the strut systems, computed tomography (CT) is used, which has been carried out by the commercial supplier YXLON Inspection Services. The provided data sets have a spatial resolution (voxel size) of $\Delta z = 115\,\mathrm{\mu m}$, which results in array sizes of 348$\times$261$\times$1521 voxels. The gray scale resolution is 8~bit or 256 gray values.

\begin{figure}
    \centering
    \begin{subfigure}[b]{0.44\textwidth}
    	\def\svgwidth{\textwidth}
    	\large
\begingroup%
  \makeatletter%
  \providecommand\color[2][]{%
    \errmessage{(Inkscape) Color is used for the text in Inkscape, but the package 'color.sty' is not loaded}%
    \renewcommand\color[2][]{}%
  }%
  \providecommand\transparent[1]{%
    \errmessage{(Inkscape) Transparency is used (non-zero) for the text in Inkscape, but the package 'transparent.sty' is not loaded}%
    \renewcommand\transparent[1]{}%
  }%
  \providecommand\rotatebox[2]{#2}%
  \newcommand*\fsize{\dimexpr\f@size pt\relax}%
  \newcommand*\lineheight[1]{\fontsize{\fsize}{#1\fsize}\selectfont}%
  \ifx\svgwidth\undefined%
    \setlength{\unitlength}{3220.00003845bp}%
    \ifx\svgscale\undefined%
      \relax%
    \else%
      \setlength{\unitlength}{\unitlength * \real{\svgscale}}%
    \fi%
  \else%
    \setlength{\unitlength}{\svgwidth}%
  \fi%
  \global\let\svgwidth\undefined%
  \global\let\svgscale\undefined%
  \makeatother%
  \begin{picture}(1,0.81397514)%
    \lineheight{1}%
    \setlength\tabcolsep{0pt}%
    \put(0,0){\includegraphics[width=\unitlength,page=1]{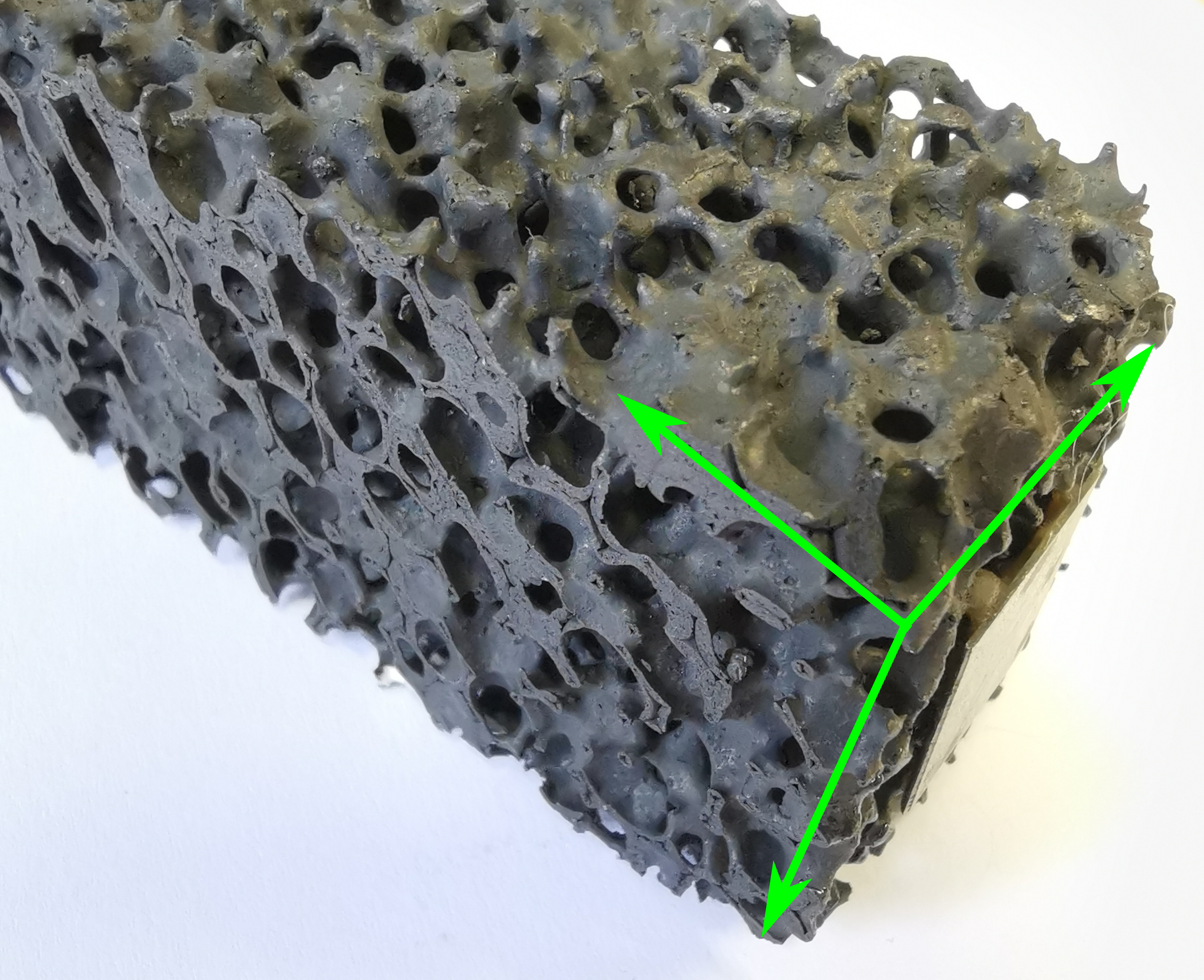}}%
    \put(0.95696539,0.4698314){\color[rgb]{0,1,0}\makebox(0,0)[lt]{\lineheight{1.25}\smash{\begin{tabular}[t]{l}x\end{tabular}}}}%
    \put(0.5780827,0.03964363){\color[rgb]{0,1,0}\makebox(0,0)[lt]{\lineheight{1.25}\smash{\begin{tabular}[t]{l}y\end{tabular}}}}%
    \put(0.46458448,0.45281739){\color[rgb]{0,1,0}\makebox(0,0)[lt]{\lineheight{1.25}\smash{\begin{tabular}[t]{l}z\end{tabular}}}}%
  \end{picture}%
\endgroup%

		\caption{real foam}\end{subfigure}\hfill
       \begin{subfigure}[b]{0.425\textwidth}
    	\def\svgwidth{\textwidth}
    	\large
\begingroup%
  \makeatletter%
  \providecommand\color[2][]{%
    \errmessage{(Inkscape) Color is used for the text in Inkscape, but the package 'color.sty' is not loaded}%
    \renewcommand\color[2][]{}%
  }%
  \providecommand\transparent[1]{%
    \errmessage{(Inkscape) Transparency is used (non-zero) for the text in Inkscape, but the package 'transparent.sty' is not loaded}%
    \renewcommand\transparent[1]{}%
  }%
  \providecommand\rotatebox[2]{#2}%
  \newcommand*\fsize{\dimexpr\f@size pt\relax}%
  \newcommand*\lineheight[1]{\fontsize{\fsize}{#1\fsize}\selectfont}%
  \ifx\svgwidth\undefined%
    \setlength{\unitlength}{786.75bp}%
    \ifx\svgscale\undefined%
      \relax%
    \else%
      \setlength{\unitlength}{\unitlength * \real{\svgscale}}%
    \fi%
  \else%
    \setlength{\unitlength}{\svgwidth}%
  \fi%
  \global\let\svgwidth\undefined%
  \global\let\svgscale\undefined%
  \makeatother%
  \begin{picture}(1,0.81982841)%
    \lineheight{1}%
    \setlength\tabcolsep{0pt}%
    \put(0,0){\includegraphics[width=\unitlength,page=1]{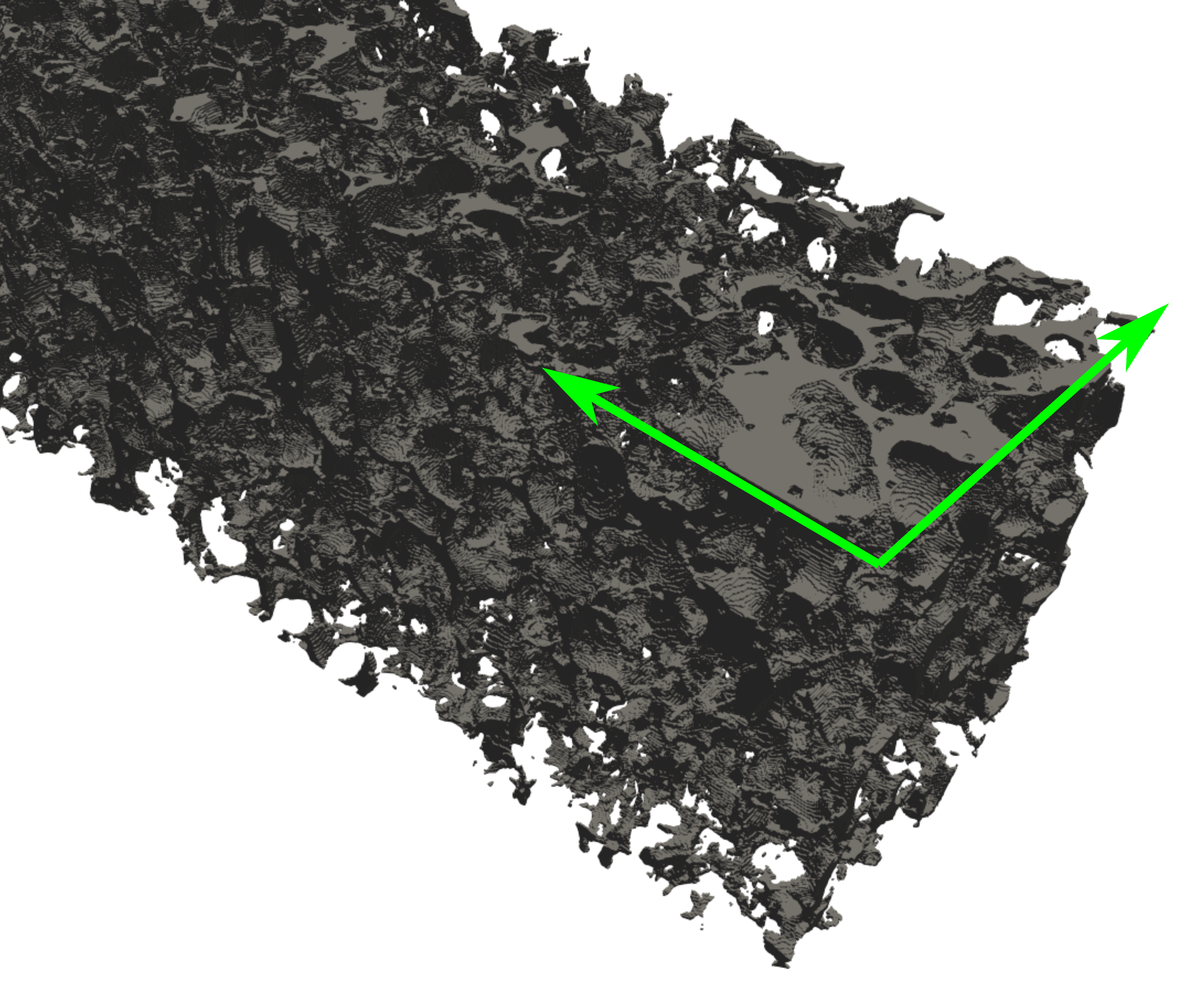}}%
    \put(0.96923107,0.49939681){\color[rgb]{0,1,0}\makebox(0,0)[lt]{\lineheight{1.25}\smash{\begin{tabular}[t]{l}x\end{tabular}}}}%
    \put(0.7114419,0.02650415){\color[rgb]{0,1,0}\makebox(0,0)[lt]{\lineheight{1.25}\smash{\begin{tabular}[t]{l}y\end{tabular}}}}%
    \put(0.42315691,0.45393702){\color[rgb]{0,1,0}\makebox(0,0)[lt]{\lineheight{1.25}\smash{\begin{tabular}[t]{l}z\end{tabular}}}}%
    \put(0,0){\includegraphics[width=\unitlength,page=2]{virtuell_Koordinatensystem.pdf}}%
  \end{picture}%
\endgroup%

		\caption{segmented CT-image}\end{subfigure}
    \caption{Comparison of a section of the real specimen with the corresponding part of the CT-scan. The highlighted red volume represents the dimensions of the smallest virtual sample.}
    \label{fig:comparison_real_ct}
\end{figure}

Subsequently, several image processing steps have been performed to remove noise in the pore areas and achieve a better contrast between fore- and background. Afterwards the strut system is identified via thresholding, where the threshold is chosen in such a way, that the relation
\begin{equation}\sum\limits_{i=1}^N c_i\cdot \rhobulk\left(\Delta z\right)^3 = \mspec\end{equation}
holds. Here, the mass $\mspec$ is the weighed mass of the corresponding specimens. This step is necessary due to the fact that in common computed tomography devices there is no one-to-one correlation between the determined gray value at a certain position and the corresponding mass. The value $c_i$ of the voxel identifies foreground (1) or background (0) at a certain position $i$ in linear indexing, where $N$ is the number of voxels in the CT scan. The real structure in comparison with the reconstruction from CT can be seen in figure~\ref{fig:comparison_real_ct}.

The pore diameters in both specimens have been extracted via image segmentation by watershed transformation \cite{Ohser.2010}. The knowledge of the pore diameters is important for the normalization of the sample height for the following evaluation. Because of the stochastic characteristic of the foam structure there occurs a distribution of the pore diameters. Further information on the distribution can be found in \cite{Kirchhof.2022}. Additionally, the pore diameters in $y$-direction are higher than in the other two spatial directions indicating a slightly anisotropic characteristic of the structure. The mean pore diameters for the specimen $\dpx$ in $x$-direction, $\dpy$ in $y$-direction, $\dpz$ in $z$-direction as well as the diameters of a corresponding ball $\dpball$ and relative density $\rhorel$ are determined as listed in table~\ref{tab:parameters_samples}. 
\begin{table}
    \centering
     \caption{Parameters of the foam specimen used for the generation of virtual samples. $\Eeff$, $\Geff$ and $\nueff$ have been determined from results of largest sample during the virtual experiments.}
    \begin{tabular}{l c|c|c}
         & &specimen 1&specimen 2 \\\hline
         $\rhorel$& [1]&0.194 &0.2\\\hline
         $\dpx$ & [mm] & 5.3 & 5.8 \\ \hline
         $\dpy$ & [mm] & 6.7 & 6.9 \\ \hline
         $\dpz$ & [mm] & 5.6 & 5.6 \\ \hline
         $\dpball$ & [mm] & 5.5 & 5.1 \\ \hline         
         $\Eeffbx$ &[GPa] & 0.433 & 0.458\\\hline
         $\Eeffby$ &[GPa] & 0.482 & 0.487\\\hline
         $\Eeffl$ &[GPa] & 0.486 & 0.579\\\hline
         $\Geff$ &[GPa] & 0.240 & 0.248\\\hline
         $\nueff$& [1] & 0.01 & -0.01\\\hline
    \end{tabular}
   
    \label{tab:parameters_samples}
\end{table}
Out of these parameters, the value $\dpball$ will be used for the following normalization of the specimen dimensions.      

The table does already contain the effective values of Young's modulus in bending (around $x$ and $y$-axes), in longitudinal loading and of shear modulus from torsion from the virtual experiments as described below.
The differences in the average values of Young's moduli determined from both bending directions and longitudinal loading of 10--15\% is not insignificant, but it is definitively smaller than the other effects which will be investigated below. 
At this point it shall only be mentioned that the effective Young's modulus of other specimens of the material has been determined also experimentally by means of vibration measurement as described in \citep{Kirchhof.2022} giving values  $\Eeffby=0.689$~GPa and $\Eeffl=0.749$~GPa. These values exceed those in table~\ref{tab:parameters_samples} by about 30\%. 
This deviation might be an indication that the $\Ebulk$ mentioned in section~\ref{sec:foammaterial} could be too low for the present charge of material.
Though, the effect of $\Ebulk$ is irrelevant for the following investigations where the results are presented in normalized form.

With the CT-scans available, samples for the virtual experiments are \enquote{cut out} from the voxel data of both specimens as follows: 
\begin{enumerate}
    \item sample size in $x$-direction is chosen as multiples of 16 voxel.
    \item sample size in $y$-direction is roughly 1.5 times the size in $x$ for better separability of the eigenmodes.
    \item size in $z$-direction is 1520 voxel.
    \item position of the sample in the $x$-$y$-plane of the CT-scan is set by two random integers which specify the origin of the coordinate system. 
\end{enumerate}
The number of generated virtual samples for different sample sizes can be found in table~\ref{tab:specimen_dimensions_and_quantity}.

\begin{table}
    \centering
     \caption{Dimensions and quantity of generated virtual samples per specimen}
    \begin{tabular}{l|c|c}
         size ($x$/$y$/$z$)&  specimen 1& specimen 2 \\\hline
         7.36 mm $\times$ 11.04 mm $\times$ 175 mm& 4 samples&6 samples \\\hline
         11.04 mm $\times$ 16.56 mm $\times$  175 mm &6 samples &6 samples \\\hline
         16.56 mm $\times$ 23.92 mm $\times$  175 mm & 4 samples &5 samples\\\hline
         23.92 mm $\times$ 36.8 mm $\times$ 175 mm & 4 samples &4 samples\\\hline
    \end{tabular}
  
    \label{tab:specimen_dimensions_and_quantity}
\end{table}

\subsection{Virtual Experiments}

The aforementioned virtual samples (as cut from the CT voxel image of the complete foam specimens) are tested virtually by fully resolved simulations using the finite cell method with the open source solver FCMLab \cite{Zander.2014}. 
Linear elastic behavior is assumed for the bulk material using $\Ebulk$ and $\nubulk$ as given in the previous section~\ref{sec:foammaterial}.
In order to avoid any effects of locally disturbed deformation fields by bearings or the kind of load application, the dynamic eigenmodes and eigenfrequencies of the free virtual samples are computed. Out of these, the angular eigenfrequencies $\eigenfreqangbendx$, $\eigenfreqangbendy$, $\eigenfreqangtors$, $\eigenfreqanglong$ of the first bending mode in each direction, the first torsional mode and the first longitudinal mode, respectively, are considered.

In conventional beam theory, these angular eigenfrequencies are
\begin{align}
    \eigenfreqangbend \approx &\frac{4.73^2}{\lspec^2}\sqrt{\frac{EI_{\mathrm{b}}}{\rho A}\cdot\frac{1+12\frac{I_\mathrm{b}}{\lspec^2 A}}{1+\pi^2\frac{I_\mathrm{b}}{\lspec^2 A}\left(\frac{E}{\kappa G}+2\pi\right)}}
    \label{eq:eigenfrequencybend} \\
    \eigenfreqangtors=& \frac{\pi}{\lspec} \sqrt{\frac{GI_{\mathrm{t}}}{\rho I_{\mathrm{p}}}}  \label{eq:eigenfrequencytors}\\
    \eigenfreqanglong=& \frac{\pi}{\lspec} \sqrt{\frac{E}{\rho}}  \label{eq:eigenfrequencylong}\,.
\end{align}
Therein, $\lspec$ refers to the length of the sample under consideration and $I_{\mathrm{b}}$, $I_{\mathrm{p}}$ and $I_{\mathrm{t}}$ denote the second moment of area with respect to bending, the polar second moment of area and the torsional moment of area of the respective cross sections. 
For a rectangular cross section $A=w\times h$ under consideration, the cross sectional properties amount to
$I_{\mathrm{b}}=\frac{wh^3}{12}$, $I_{\mathrm{p}}=wh\frac{h^2+w^2}{12}$, and, for $h/w=1.5$, $I_{\mathrm{t}}=0.196hw^3$.
Note that Eq.~\eqref{eq:eigenfrequencybend} even includes the correction from the Timoshenko beam theory \cite{Shi.2016} in order to avoid any artefacts from the slenderness ratio of the samples (although this term does not affect the principal trends shown in the following). It holds $\kappa=5/6$ while the ratio of elastic modulus and shear modulus can be evaluated as $E/G=2(1+\nu)$. For the evaluation, a constant value $\nu=\nueff$ according to the values in table~\ref{tab:parameters_samples} is assumed (having in mind that $\nu$ has a negligible influence in Eq.~\eqref{eq:eigenfrequencybend} at all).

Having computed the respective angular eigenfrequencies from the virtual tests, the \emph{apparent values} $\Gapp$ and $\Eapp$ of shear modulus and Young's modulus, respectively, are obtained by solving Eqs.~\eqref{eq:eigenfrequencybend}--\eqref{eq:eigenfrequencylong}. The values $\Eappbend$ and $\Eapplong$ of the apparent Young's modulus are marked with an additional index to indicate whether they have been extracted from the first bending mode using Eq.~\eqref{eq:eigenfrequencybend} or from the longitudinal mode using Eq.~\eqref{eq:eigenfrequencylong}, respectively.

\section{Results}

\begin{figure}
    \centering
    \begin{subfigure}[b]{0.51\textwidth}
        \def\svgwidth{\textwidth}
        \scriptsize
    	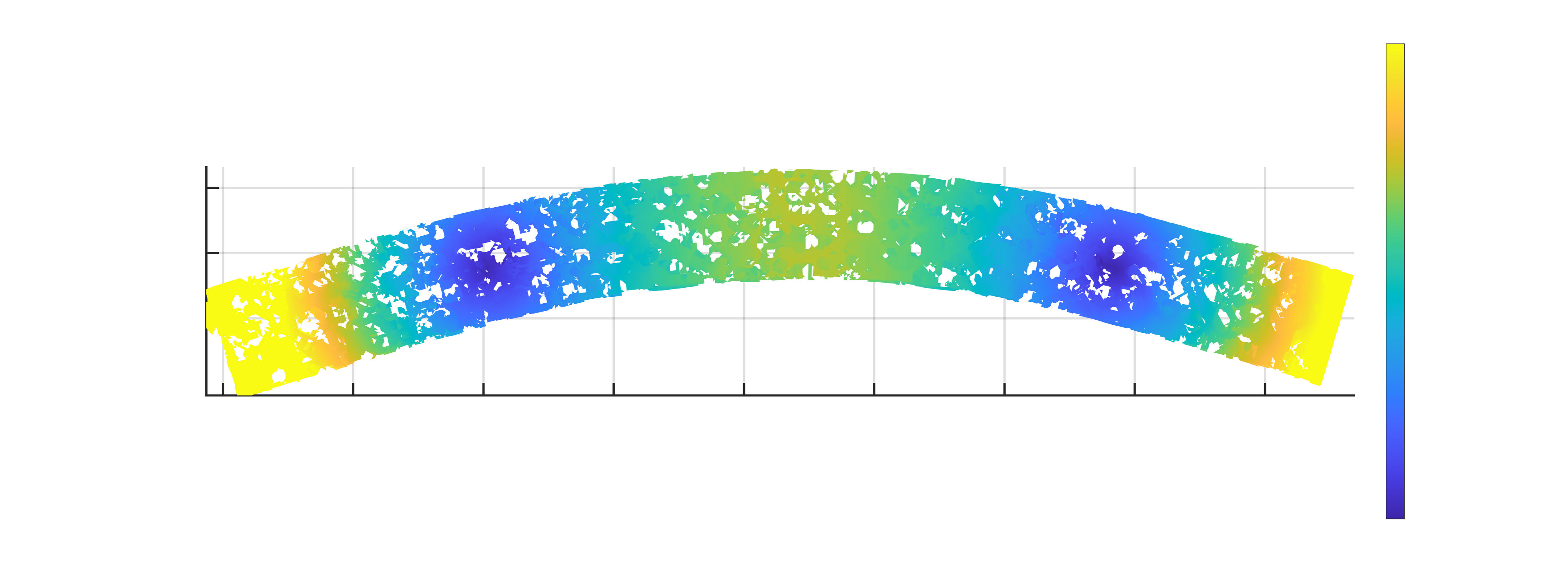
            \caption{Bending}
        \label{fig:eigenmodes_bending}    
    \end{subfigure}\hfill
    \begin{subfigure}[b]{0.48\textwidth}
        \def\svgwidth{1\textwidth}
        \scriptsize
    	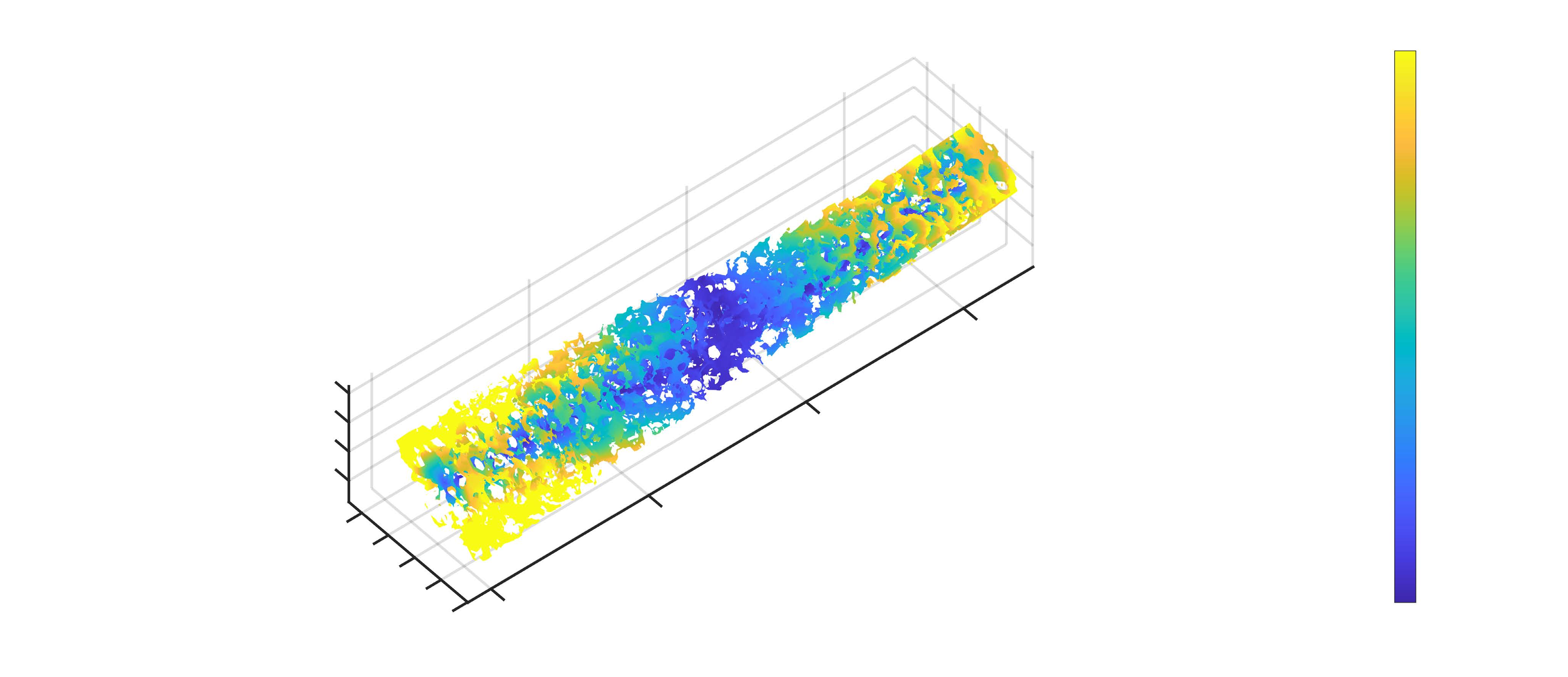
        \caption{Torsion}
        \label{fig:eigenmodes_torsion}
    \end{subfigure}
    \caption{Eigenmodes from virtual experiments}
    \label{fig:eigenmodes}
\end{figure}

Figure~\ref{fig:eigenmodes} shows, exemplarily, two eigenmodes from the virtual tests, which have been used to extract the apparent values of Young's modulus and shear modulus under bending and torsion, respectively.
In the following, the results are represented in terms of the size effect ratio \citep{Anderson1994a}
\begin{equation}
    \sizeeffratio=\frac{\Eapp}{\Eeff}
    \label{eq:sizeeffectratio}
\end{equation}
as ratio of the measured (real or virtual) response and the respective prediction of classical Cauchy theory of continuum mechanics. In particular, Eq.~\eqref{eq:sizeeffectratio} is formulated in terms of Young's modulus, whereby $\Eeff$ is the effective value which would be measured at a theoretically infinitely large specimens. 
Practically, the effective Young's modulus $\Eeff$ is taken as average of $\Eapp$ of the largest samples within each testing series as listed in table~\ref{tab:parameters_samples}. This condition means that the size effect ratio $\sizeeffratio$ of the largest samples equals unity (within reasonable range of accuracy).
Using $\sizeeffratio$ instead of absolute values of Young's modulus has the advantage that Young's modulus $\Ebulk$ of the bulk material of the foam drops out so that foams with different bulk materials can be compared (and actually, $\Ebulk$ is just an input value to the present virtual experiments).

Again, the indices of the size effect ratios $\sizeeffratiobend$, $\sizeeffratiolong$ and $\sizeeffratiotors$ indicate whether they are determined for bending, longitudinal deformation or torsion, respectively, whereby $\sizeeffratiotors$ is of course defined with respect to shear modulus $\Gapp$ instead of Young's modulus.

\subsection{Bending}

\begin{figure}
    \centering
    \begin{subfigure}{0.46\textwidth}
        \includegraphics[width=\textwidth]{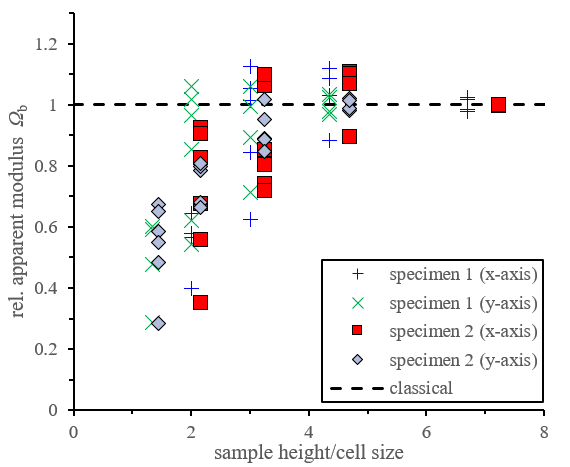}
        \caption{bending}
        \label{fig:sizeeffectbend}
    \end{subfigure}\hfill
    \begin{subfigure}{0.46\textwidth}
        \includegraphics[width=\textwidth]{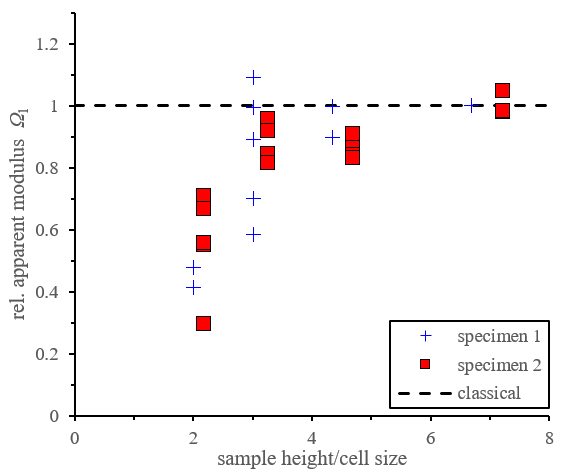}
        \caption{longitudinal}
        \label{fig:sizeeffectlong}
    \end{subfigure}
    \begin{subfigure}{0.46\textwidth}
        \includegraphics[width=\textwidth]{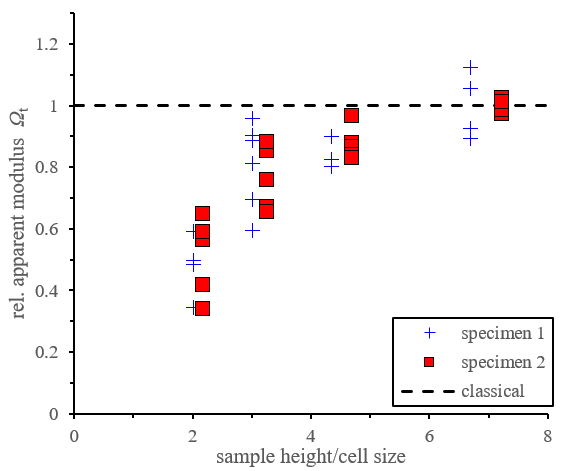}
        \caption{torsion}
    \label{fig:sizeeffecttors}
    \end{subfigure}
    \hfill
    \begin{subfigure}{0.46\textwidth}
        \includegraphics[width=\textwidth]{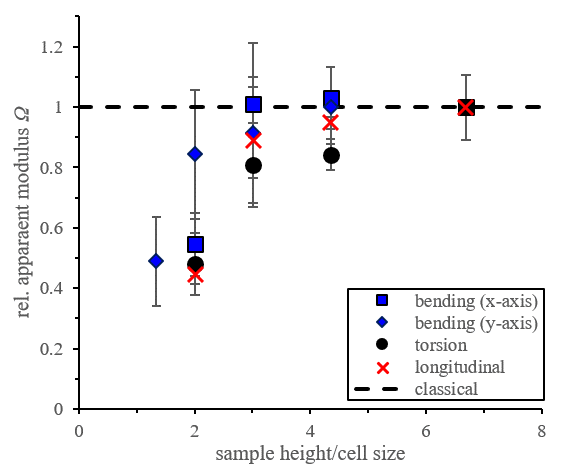}
        \caption{comparison}
        \label{fig:sizeeffect_tors_long_cmp}
    \end{subfigure}
    \caption{Relative apparent moduli for different loading modes}
\end{figure}

Figure~\ref{fig:sizeeffectbend} shows a plot of the extracted values of the apparent Young's modulus of different virtual samples in normalized form of $\sizeeffratio$ versus the relative sample size. The latter is defined as absolute sample size (here the height is taken) divided by $\dpball$ of the specimen under investigation.
First of all, it can be said that the mean values of the two largest sample sizes of each specimen lie close to unity. Thus, it can be concluded that they were large enough to identify the effective Young's modulus $\Eeff$ with reasonable accuracy.  
Secondly, figure~\ref{fig:sizeeffectbend} shows a clear negative size effect, i.e., specimens with a ratio of specimen height and cell size below about 3 or 4 show a lower apparent Young's modulus $\Eapp$ than larger specimens, so that the average size effect ratio $\sizeeffratio$ lies below unity in this regime. 
Obviously, the \emph{edge softening effect} of less connected struts near the specimen surface \citep{Brezny1990} seems to be much more relevant, at least for the present foam, than the additionally transmitted moments of the struts \citep{Anderson1994a}. For the first time it can now concluded that this edge softening effect is \emph{no artefact} of the cutting or machining of the specimens as this was done virtually in the present study with ideal precision. 

Regarding the bending around $x$-axis and $y$-axis, there seems to be at least no significant anisotropy as it could appear, e.g., if the main principal axes of the pores would be rotated against the chosen coordinate system.

Furthermore, figure~\ref{fig:sizeeffectbend} shows that the scatter of $\Eapp$ and thus of $\sizeeffratio$ increases with decreasing specimen size. This behavior is plausible and expected since smaller specimens contain less pores in direction of  thickness, so that the relative influence of single pores and their particular realization does increase within this regime.

\subsection{Uniaxial deformation}

Figure~\ref{fig:sizeeffectlong} presents respective data for the longitudinal modes of deformations, being associated with a uniaxial stress state in classical theory of rods.
In this case, there is only a single relevant mode for each cut sample of the two specimens, so that only one graph per specimen is obtained.
Qualitatively, the picture in figure~\ref{fig:sizeeffectlong} is identical to the bending case.
Smaller specimens show a smaller apparent Young's modulus than it would be expected from sufficiently large specimens.
However, a minor quantitative difference to the bending case can be observed. In particular for specimen 2, the second largest specimen with about 3.5 pores over height (in average) deviates by about 10\% from the main (thought) graph of regression.
Anyway, the latter value will presumably not differ by more than a few percent from the presently extracted values and a clear negative size effect is obvious.

\subsection{Torsion}

As third class of vibration modes with basic interpretation, the eigenfrequencies of the torsional modes, as shown in figure~\ref{fig:eigenmodes_torsion}, have been extracted and have been used to extract the apparent shear modulus $\Gapp$ via Eq.~\eqref{eq:eigenfrequencytors}.
Figure~\ref{fig:sizeeffecttors} presents these values in the same normalized way via $\sizeeffratiotors$ vs.\ cross section divided by cell size. 

Again there is only one graph per specimen because there is only one torsional mode for each sample of the specimen. Qualitatively the torsional case shows the same behavior as observed in figures \ref{fig:sizeeffectbend} and \ref{fig:sizeeffectlong}. Additionally, a greater decrease of the relative stiffness compared to the bending case can be observed while there is no quantitatively observable difference in the determined relative stiffness compared to the uniaxial case. Other than in the longitudinal case also no deviations from the thought regression graph can be noticed. Again, the difference in the apparent moduli of the two largest sample sizes of each specimen differ by about 15\%, so that it is possible again, that the effective shear modulus has not been reached completely yet with the largest samples. 
But without any doubt, a clear negative size effect is observed for torsional vibration.   

\subsection{Comparison}

The normalized apparent moduli for the three basic deformation modes are compared in figure~\ref{fig:sizeeffect_tors_long_cmp}.
The error bars correspond estimates of the standard deviation from the investigated samples of same size.

As could have been anticipated from the other three plots, the data points for the three modes lie within a common band.
It indicates a decrease of the apparent modulus if the relevant cross section of a specimen encompasses less than about 6 to 8 cells in the relevant direction.
This finding is an indication that the respective mesoscale mechanisms of incomplete cells at the surface are identical for the three modes.

\section{Discussion}

\begin{figure}
    \centering
    \begin{subfigure}{0.46\textwidth}
        \includegraphics[width=\textwidth]{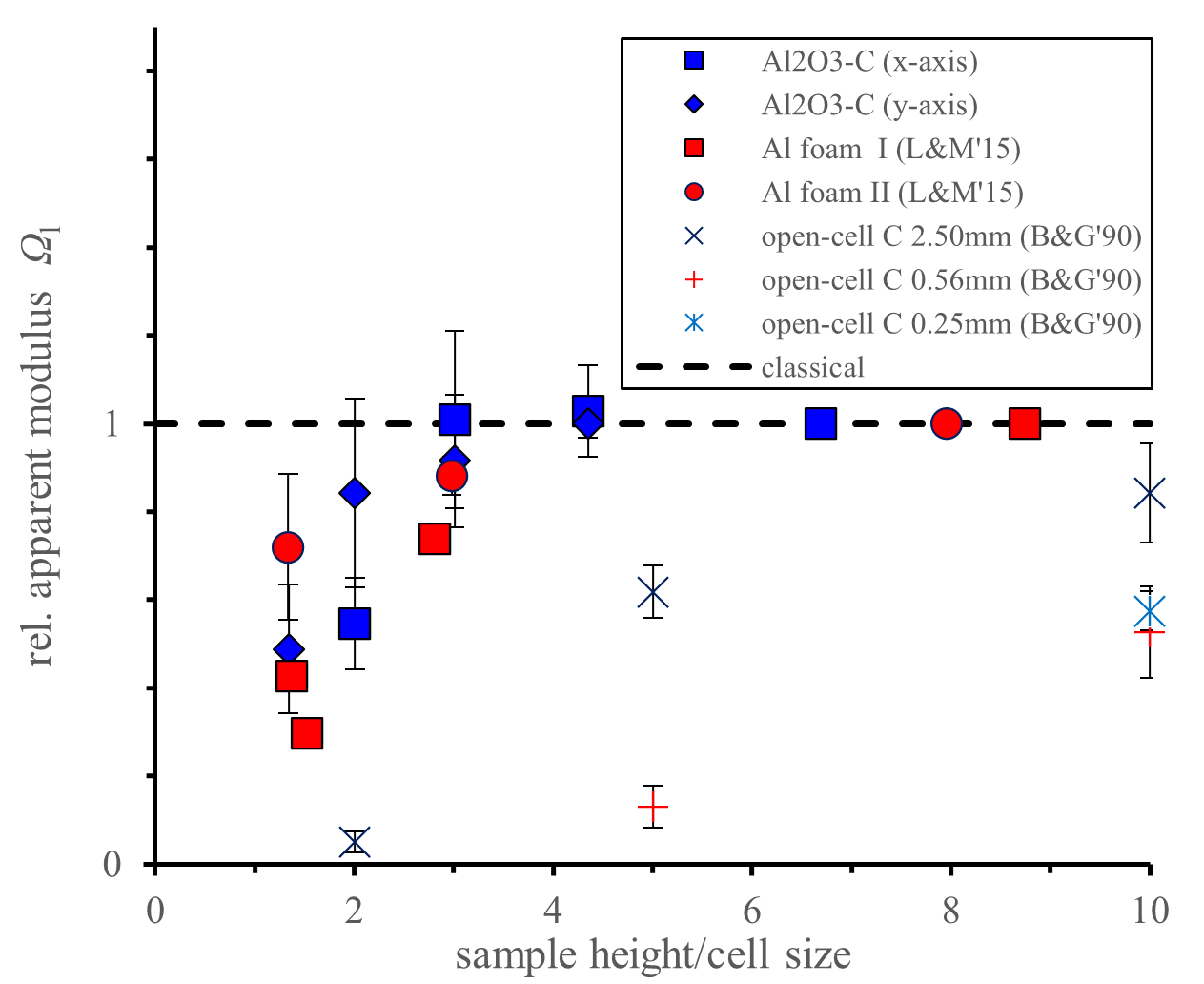}
        \caption{with negative size effect}
        \label{fig:sizeeffect_bend_cmp_neg}
    \end{subfigure}
    \hfill
    \begin{subfigure}{0.46\textwidth}
        \includegraphics[width=\textwidth]{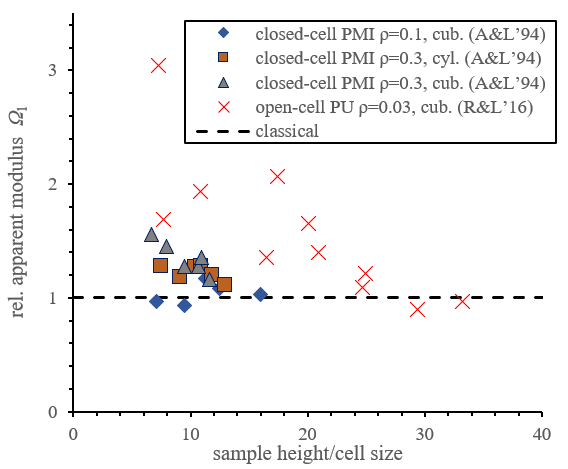}
        \caption{with positive size effect}
        \label{fig:sizeeffect_bend_cmp_pos}
    \end{subfigure}
    \caption{Comparison of different experimental results for size effects in bending from literature}
    \label{fig:sizeeffect_bend_cmp}
\end{figure}

In the next step, the results of the present study are compared to experimental results from literature for size effects in the elastic behavior of foams, starting with bending in figure~\ref{fig:sizeeffect_bend_cmp}.

Figure~\ref{fig:sizeeffect_bend_cmp_neg} comprises the data of \citet{Brezny1990} (abbreviated \enquote{B\&G'90} in the legend) and of Liebold and Müller \citep{Liebold2015,Liebold2016a} (abbreviated \enquote{L\&M'15}) who had observed negative size effects as well.
\citeauthor{Brezny1990} had investigated low-density foams ($\rhorel\approx0.035$) of glassy carbon from converted open-cell polymer foams. Liebold and Müller obtained the apparent moduli of aluminum foams from two production routes (foam-I from powder metallurgical production and foam-II produced by gas injection method) by vibrational analysis.
From the given mass density compared to the one of compact aluminum, the relative density $\rhorel$ can be estimated to be $0.15$ and $0.10$, respectively.

It can be observed from figure~\ref{fig:sizeeffect_bend_cmp_neg} that the present normalized results do match even quantitatively with those of Liebold and Müller within a band of reasonable accuracy. In contrast, the data of \citeauthor{Brezny1990} differ significantly from this band. Even the three chemically identical foams with different pore sizes (2.5~mm, 0.56~mm, 0.25~mm) of \citeauthor{Brezny1990} differ significantly among each other, even in the normalized representation. That is why  \citeauthor{Brezny1990} argued that artefacts from the manufacturing process could be relevant (and actually their intention was to identify a sufficient specimen size to measure $\Eeff$ reliably).

In contrast, Lakes and co-workers \citep{Anderson1994a,Rueger2016} had observed a positive size effect in bending. The results of their studies are compared quantitatively in figure~\ref{fig:sizeeffect_bend_cmp_pos}.

Figures~\ref{fig:sizeeffect_bend_cmp_neg} and \ref{fig:sizeeffect_bend_cmp_pos} do not only show the opposite trend, but regarding the abscissa of both plots it can even be found that the size effect was observed within different regions of scale separation ratio height/cell size, in particular for the low-density open-cell PU foam\footnote{A second foam of the same type but with cell size 0.4~mm was tested in \citep{Rueger2016} as well. But in the normalization w.r.t.\ cell size, the relevant data points thereof do even lie outside the given horizontal axis of figure~\ref{fig:sizeeffect_bend_cmp_pos}. This applies to the data from \citep{Lakes1983,Lakes1986} as well.} (cell size: 1.2~mm) investigated by \citet{Rueger2016} (abbreviated \enquote{R\&L'16}). 
Though, the low-density PMI foam $\rhorel=0.1$ from the study of \citet{Anderson1994a} (\enquote{A\&L'94}) does at least not contradict the data of Liebold and Müller and of the present study in figure~\ref{fig:sizeeffect_bend_cmp_neg}.

\begin{figure}
    \centering
    \begin{subfigure}{0.46\textwidth}
    \centering
    \includegraphics[width=\textwidth]{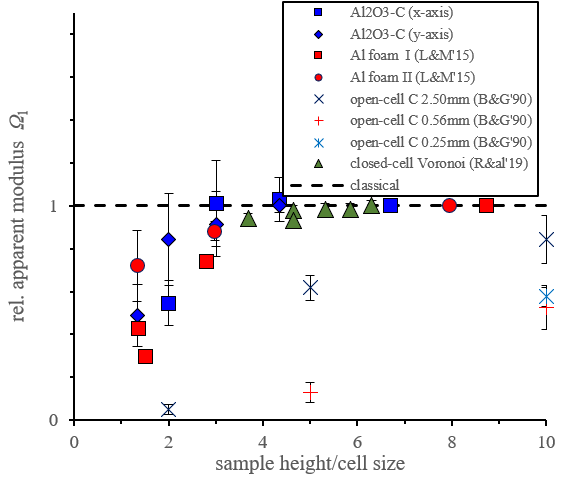}
    \caption{uniaxial compression}
    \label{fig:sizeeffect_long_cmp}
    \end{subfigure}
    \hfill
    \begin{subfigure}{0.46\textwidth}
        \includegraphics[width=\textwidth]{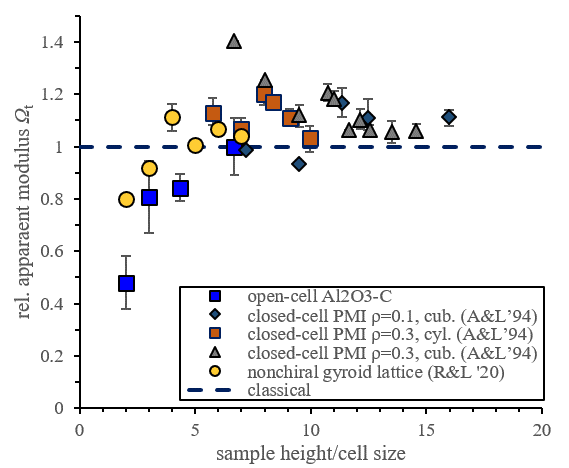}
        \caption{torsion}
        \label{fig:sizeeffect_torsion_cmp}
    \end{subfigure}
    \caption{Comparison of different experimental results from literature with present simulation results}
\end{figure}

Figure~\ref{fig:sizeeffect_long_cmp} compares the results of the present study for the longitudinal mode with the experimental data from compression tests by Gibson, Onck and co-workers \citep{Tekoglu2011,Andrews2001} (\enquote{G\&O\&al}) and by \citet{Rueger2016}.
Gibson, Onck and co-workers investigated two commercial aluminum foams (estimated $\rhorel=0.07\ldots 0.08$) with different cell topologies. The normalized compression data of both of them comply very well with the results of the present study.
In addition, figure~\ref{fig:sizeeffect_long_cmp} comprises compression data from the study of \citet{Rueger2016} which has been already considered for bending. Due to the large scatter of their data points, the effective modulus $\Eeff$ and thus the resulting level $\sizeeffratiolong=1$ in the plot is taken here as the average of the 6 largest samples (out of 12 in total).
Qualitatively, the measurements of \citet{Rueger2016} exhibit a negative size effect in compression as well. But the scatter within the relatively low number of data points is too large to allow for a sound quantitative comparison.
Furthermore, figure~\ref{fig:sizeeffect_long_cmp} shows the results of 3d direct numerical simulations (DNS) of the stochastic mesostructures of a closed-cell foam by \citet{Rajput2019} (\enquote{R\&al'19}) generated by Voronoi tesselation. These data lie in the common band of the present study and the aforementioned ones. This applies also to the 2d stochastic DNS with beam elements of \citet{Tekoglu2011} (not shown here).

Figure~\ref{fig:sizeeffect_torsion_cmp} shows the present results for torsion in comparison to respective data for two PMI foams of \citet{Anderson1994a} (\enquote{A\&L '94}) and additively manufactured nonchiral gyroid structure from \citet{Reasa2020} (\enquote{R\&L '20}). The latter exhibit a region with negative size effect below a ratio of sample height and cell size of about 4 or 5 as the present data, the results for the lighter PMI foam with $\rhorel=0.1$ are at least not contradictory to such a trend. Only the denser PMI foam with relative density $\rhorel=0.3$ exhibits a clear positive size effect. Very recently, \citet{Reasa2020} formulated the hypothesis that there might be regimes regarding sample size with opposite size effects. The present data does not show these regimes, but their presence cannot be excluded until larger samples were considered.

\section{Summary and Conclusions}

Within the present study, 39 virtual samples of different sizes were cut from CT scans of real open-cell ceramic foams. 
The apparent elastic moduli of these samples were determined by means of free vibrational analysis.
For this purpose, eigenfrequencies for bending, torsion and uniaxial loading have been computed using an open-source finite cell code.
This procedure ensures that the apparent moduli are neither influenced by the sample preparation process nor by disturbances from load application or bearings.
All of the three basic modes exhibit a clear negative size effect. Smaller specimens are more compliant than it would be expected from larger ones when using classical theory of elasticity.

In the second part of the present paper, the obtained results have been compared to (the relatively few available) experimental data from literature for foams.
In uniaxial loading, the present study complies with all found data in literature qualitatively and even quantitatively to a certain extent, if the data is normalized to cell size.
Such a quantitative agreement was also found with respect to apparent Young's modulus in bending with respective to another study of Liebold and Müller and qualitatively with one of \citeauthor{Brezny1990}.
However, Lakes and co-workers had observed the opposite trend within a number of studies in bending as well as in torsion.
These discrepancies could not be attributed to common basic parameters of foams like relative density or cell topology (closed-cell or open-cell). Therefore representatives of each category with different relative densities are among those with positive or negative size effect in literature.
Recently, \citet{Reasa2020} observed two regimes of sample sizes with opposite trends for gyroid lattices.

It must be concluded, that the present state of knowledge on elastic size effects for real foams in bending and torsion seems to be incomplete. Further careful experiments are required. 
In particular, the large scatter of single data points within the relevant regime of small samples has to be taken into account via a sufficient number of test and/or by repetition from independent research groups.
The proposed method of virtual testing of cuts from CT images offers a way to do so in a reliable way and with less effort than with real samples.

\section{Acknowledgement}
The authors want to thank the Instiute of Ceramics, Refractories and Composite Materials of the TU Bergakademie Freiberg for the provision of sample material.
\bibliographystyle{plainnat}
\bibliography{references}
\end{document}